\documentclass[%
 reprint,
superscriptaddress,
 nofootinbib,
 amsmath,amssymb,
aps,
prl,
 showkeys,
]{revtex4-2}

\usepackage{ruths_commands}
\usepackage{martins_commands}

\begin{document}

\title{A universal scaling law for gravitational waves induced during inflation}

\author{Martin Teuscher}
\email{teuscher@lpsc.in2p3.fr}

\affiliation{Laboratoire de Physique Subatomique et de Cosmologie\char`,{} Univ. Grenoble-Alpes\char`,{} CNRS-IN2P3\char`,{} 53 avenue des Martyrs\char`,{} 38000 Grenoble\char`,{} France}
\affiliation{Département de Physique Théorique\char`,{} Université de Genève\char`,{}  24 quai Ernest Ansermet\char`,{} 1211 Genève 4\char`,{} Switzerland}

\author{Ruth Durrer}
\email{ruth.durrer@unige.ch}

\affiliation{Département de Physique Théorique\char`,{} Université de Genève\char`,{}  24 quai Ernest Ansermet\char`,{} 1211 Genève 4\char`,{} Switzerland}

\author{Killian Martineau}
\email{martineau@lpsc.in2p3.fr}
\author{Aurélien Barrau}
\email{barrau@in2p3.fr}

\affiliation{Laboratoire de Physique Subatomique et de Cosmologie\char`,{} Univ. Grenoble-Alpes\char`,{} CNRS-IN2P3\char`,{} 53 avenue des Martyrs\char`,{} 38000 Grenoble\char`,{} France}

\date{\today}


\begin{abstract}

We consider the stochastic gravitational wave background induced by arbitrary source fields that are amplified during cosmological inflation. 
The associated tensor spectral index is shown to be given, under minimal assumptions, by a simple formula easy to apply in most situations of accelerated expansion. For slow-roll inflation, the induced spectrum is nearly scale invariant, with an index deviating from the standard outcome of vacuum generated gravitational waves. Remarkably, we demonstrate that scale invariance remains true regardless of the original spectrum of the source. We show how this generic approach reproduces the literature on specific models of gravitational wave primordial sources, and discuss its limitations. It provides a very practical estimation of the tensor spectral index for future models, to which subleading corrections can then be added.
\end{abstract}


\maketitle

\section{Introduction}
\label{sec:intro}


Cosmological stochastic gravitational wave backgrounds (SGWBs) originating from inflation have attracted considerable attention \cite{Guzzetti:2016mkm}, as one of the few signatures that may be used to  directly probe the high-energy early Universe \cite{Caprini:2018mtu}. It is well known that a period of inflation leads to the amplification of quantum fluctuations of all fields that are not conformally coupled. The enhancement of a mode occurs whenever its wavelength grows larger than the Hubble  radius\footnote{As during inflation the Hubble radius is strictly speaking not a `horizon', we call the corresponding modes `super-Hubble' or `sub-Hubble', as opposed to the often used but misleading terminology of `super-horizon' or `sub-horizon'.}. Gravitational waves (GWs) are no exception to this process, which results in the formation of a nearly scale invariant stochastic background~\cite{Starobinsky:1979ty}. 

Yet, a distinct contribution to SGWBs can arise from the presence of additional stochastic fields in the Universe, whose energy-momentum tensors usually have nonvanishing transverse-traceless components that classically source GWs \cite{maggiore_vol1, Durrer:2020fza, Speziale:2025zjp}. This driven production of GWs has been studied during inflation, including GWs induced by second-order scalar perturbations \cite{Baumann:2007zm, Pi:2020otn, Domenech:2021ztg} or by  gauge fields \cite{Barnaby:2011vw,Caprini:2014mja,Teuscher:2025jhq,vonEckardstein:2025oic, vonEckardstein:2025elq}, as well as during later cosmological stages (see Ref. \cite{Caprini:2018mtu} for a review or Ref. \cite{Cai:2019cdl} for a generic result). Interestingly, these induced GWs may actually constitute the dominant contribution to the inflationary SGWB \cite{Caprini:2018mtu,Teuscher:2025jhq}.

In this work, we demonstrate a general result on the spectral index $n_T$ of such SGWBs induced during inflation. We show that our statements hold under very minimal and realistic assumptions on the nature of the source, supported by dimensionality arguments. Our main finding is {\it a very concise formula providing the GW spectral index as a function of the spectral index of the source. In particular, during slow-roll inflation, the induced GW spectrum is always nearly scale invariant, irrespective of the source power spectrum. This holds as long as the latter is not red.} 

The expression we derive for $n_T$ is valid for accelerated expansion models, parametrized by a constant equation of state $w=P/\rho < -1/3$. Even if the equation of state is not entirely constant, our approximation holds if it is slowly varying during the time when the wavelengths considered exit the Hubble scale. Very generically, the anisotropic stress tensor responsible for GW production is quadratic in the source field \cite{Auclair:2022jod,Cai:2019cdl, Domenech:2021ztg}, 
\begin{equation}
\label{eq:quadratic-Pi}
    \Pi_{ij}\propto \Psi_i\Psi_j \mperiod
\end{equation}
The precise link between $\Psi_i$ and the genuine physical field
will be elucidated below. In a wide range of scenarios, the power spectrum of $\Psi_i$ follows a power law with respect to wave mode $k$, $\PP_{\Psi} \propto k^n$ (see, e.g., Refs. \cite{Caprini:2003vc, Caprini:2018mtu}). In this case, whenever $n\geqslant 0$, we prove the GW spectral index $n_T = \dd \ln\PP_T / \dd\ln k$ to be in the super-Hubble limit
\begin{equation}
 \label{eq:nT-main-formula}
    n_T = 2(n-2)(q+1)\mcomma
\end{equation}
 where $q=2/(1+3w) < 0$ gives the scale factor evolution with conformal time, $a\propto |\tau|^q$. We further show that under the same assumptions $\PP_T$ is frozen on super-Hubble scales.  This prediction for $n_T$ matches with previously worked out specific examples, as detailed below. We also discuss the case $n<0$ for which an infrared cutoff is required to ensure a finite anisotropic stress. A natural infrared cutoff is the beginning of inflation, in which case the resulting GW amplitude becomes sensitive to the total duration of inflation. 

The physical implications of relation \eqref{eq:nT-main-formula} are threefold. 

First, (near) scale invariance can be achieved either by slow-roll $w\simeq -1$, surprisingly independent of the source spectrum, or independently of $w$ for a source with a blue spectrum such that $n=2$ (see examples in the text). Model-dependent corrections to \Eq \eqref{eq:nT-main-formula} that do not spoil scale invariance can be expected, as we detail below.

Second, if $-1< w<-1/3$ then $q+1<0$. As a consequence, blue sources with $n>2$ tend to generate red GW spectra, and vice versa. 

Lastly, to first order in slow-roll, $q=-1-\epsilon$ with $\epsilon\ll 1$ \cite{Baumann_2022}, implying that $\PP_T \propto k^{-2(n-2)\epsilon}$. Interestingly, for $n\neq 3$, this exponent differs from the one of quantum amplified GWs, $\PP_T \propto k^{-2\epsilon}$ \cite{maggiore_vol2}, meaning both contributions can in principle be distinguished. For instance, this happens to be the case for axion inflation with the axion helically coupled to a $\mathrm{U}(1)$ gauge field source, which has $n=4$ \cite{Teuscher:2025jhq}. 


\bigskip

\noindent {\it Notations.}\quad We consider a flat FLRW background metric
\begin{equation}
\label{eq:bg-metric}
    \dd s^2 = a^2(\tau)(-\dd\tau^2+\delta_{ij}\dd x^i \dd x^j)\mperiod
\end{equation}
Derivatives with respect to conformal time $\tau<0$ (resp. physical time $t$) are denoted with a prime (resp. overdot), and we define the corresponding Hubble parameters $\HH=a'/a$ and $H = \dot{a}/a = \HH/a$. 
We call inflation a period with $\ddot a>0$, or equivalently $w<-1/3$, i.e. $q=2/(1+3w)<0$. The first slow-roll parameter is $\epsilon = - \dot{H}/H^2$.

We set $\hbar=c=1$, $\planckmass=1/\sqrt{8\pi G}$ the reduced Planck mass, and $k=a k\supsc{phys}$ the comoving momentum.  We define the unequal time two-point function of any Fourier space tensor $X_J(\bm{k})$ -- $J$ collectively labeling appropriate Lorentz indices -- as
\begin{equation}
\label{eq:correlation-def}
    \left\langle{\textstyle\sum_J}\, X_J(\bm{k},\tau) X_J^*(\bm{p},\tau')\right\rangle = (2\pi)^3 P_X(k,\tau,\tau')\delta^{(3)}(\bm{k}-\bm{p})\mperiod
\end{equation}
The dimensionless\footnote{We use `dimensionless' as an abuse of speech, meaning that $\PP_X$ has the same dimension  as $X^2$ in real space.} power spectrum $\PP_X$ is then
\begin{equation}
    \label{eq:pow-spec-def}
    \PP_X(k,\tau) = \frac{k^3}{2\pi^2}P_X(k,\tau,\tau)\mperiod
\end{equation}

\section{Sourced stochastic gravitational waves}
\label{sec:GW-basics}

We consider tensor perturbations $h_{\mu\nu}$ of the metric \eqref{eq:bg-metric}, $g_{\mu\nu} = a^2(\eta_{\mu\nu}+h_{\mu\nu})$, that are sourced by a generic anisotropic stress-energy tensor $\Pi^\mu_\nu$.  Let $h_{ij}$, $\Pi_{ij}$ be the transverse-traceless (TT) spin-2 perturbations\footnote{As this will matter for dimensional arguments, we emphasize that we use the convention of moving indices of perturbative quantities with the Kronecker delta, so that, e.g., $\Pi_{ij}=\Pi^i_j$, $\Psi_i = \Psi^i$. We take that tensors of rank 2 with one lower and one upper index are direct components of their fully Lorentz covariant equivalent, $\Pi^i_j \subset \Pi^\mu_\nu$, while others are not, $\Pi_{ij} \not\subset \Pi_{\mu\nu}$.} of the metric and of $\Pi^\mu_\nu$ (carrying implicit TT labels). They satisfy the GW equation of motion \cite{maggiore_vol2}
\begin{equation}
\label{eq:GW-eq}
   (ah)_{ij}''  + \left(k^2-\frac{a''}{a}\right)(a h)_{ij} = \frac{2}{\planckmass^2}a^3 \Pi_{ij}(k)\mperiod
\end{equation}
Rewriting \Eq\eqref{eq:GW-eq} in terms of the auxiliary time variable $x\equiv k\tau<0$, this equation admits a Green's function $G(x,y)$ that provides a solution for $h_{ij}(k,\tau)$, leading to the following power spectrum  of gravitational waves
\begin{align}
\nonumber
 \dps \PP_T &= \PP_h(k,\tau)\\
  \nonumber
 &=\frac{2}{\pi^2 \planckmass^4}\frac{1}{k a^2} \int^x_{x_i}\int^x_{x_i} G(x,y)G(x,z) \\
   \label{eq:PT-formula}
     & \hspace{1.5cm}\dps \times a^3(y) a^3(z)P_\Pi\left(k,\frac{y}{k},\frac{z}{k}\right)\dd y\dd z  \mperiod
\end{align} 

The value of $x$ at the time when the source starts to generate perturbations, $x_i = k \tau_i$, will play a crucial role in what follows. Below, we will be interested in finding a general expression for the unequal time correlator of the source, $P_\Pi$. However, already at this stage its expression cannot be completely arbitrary: the function $K(\tau_1,\tau_2) = P_\Pi(k,\tau_1,\tau_2)$ must be a {\it positive kernel} in order to ensure that $\PP_T$ is  positive \cite{Auclair:2022jod,Caprini:2007xq}. 

For a constant equation of state $w$, the Green function is given by
\begin{align}
\label{eq:full-green-func}
    G(x,z) &= xz(j_\nu(z)y_\nu(x)-y_\nu(z)j_\nu(x))\\
    \label{eq:mildly-approx-green-func}
    &\simeq x^{1+\nu} F_\nu(z)\qquad (\left|x\right| \ll 1) \\
    \label{eq:approx-green-func}
&\simeq \left(\frac{x}{z}\right)^\nu \frac{x}{1+2\nu} \qquad (\left|x\right| <\left|z\right| \ll 1)\mperiod 
\end{align}
Here $j_\nu$ and $y_\nu$ are the spherical Bessel functions of order $\nu\equiv q-1 <-1$. As will appear below, the scaling $G(x,z)\propto x^{1+\nu}$ in \Eq \eqref{eq:mildly-approx-green-func} is sufficient to derive the expression \eqref{eq:nT-main-formula} for the tensor spectral index. The exact expression of $F_\nu(z) \equiv \frac{z}{2^{1+\nu}\sqrt{\pi}}\left[\cos(\nu\pi)\Gamma\left(\frac{1}{2}-\nu\right)y_\nu(z)-\sin(\nu\pi)\Gamma\left(-\frac{1}{2}-\nu\right)j_\nu(z)\right]$ is mostly irrelevant to our purpose, and for the sake of simplicity we will use \Eq \eqref{eq:approx-green-func} throughout.  We come back to the effect of this approximation on the final value of $\PP_T$ after establishing our main results.

\bigskip 

Although our main analysis focuses on the primordial GW power spectrum, we conclude this section by describing a compact formalism to compute the GW transfer function from the end of inflation to today. We also provide more details on this analytical recipe in Ref.~\cite{Teuscher:2025jhq}. This transfer function can of course otherwise be determined numerically.

Let us assume that, following inflation, the Universe evolves through a sequence of $n$ different eras of approximately constant equations of states $w_1,\dots,w_n$, with almost instantaneous transitions. We denote by $\HH_{i|i+1}$ the conformal Hubble parameter at each of these transitions and by $\HH\ssend$ its value at the end of inflation. We define a set of $2\times2$ transition matrices $T_i$ that help match $h_{\mu\nu},h'_{\mu\nu}$ through each transition,
\begin{widetext}
\begin{align}
\label{eq:transition-matrix}
     T_i & \equiv M_{i+1}[r_{i+1}]^{-1}M_i[\ell_i] \qquad\text{with}\qquad 
M_i[x] \equiv \begin{pmatrix}
    x j_{\nu_i}(x) & x y_{\nu_i}(x) \\
    (\nu_i+1)j_{\nu_i}(x)- x j_{\nu_i+1}(x) &  (\nu_i+1)y_{\nu_i}(x)- x y_{\nu_i+1}(x) 
\end{pmatrix} \mcomma \\
\nu_i &\equiv -1+\frac{2}{(1+3w_i)} \; \forall\, 1\leqslant i \leqslant n \mcomma \qquad
\ell_i \equiv \frac{2k}{\HH_{i|i+1} (1+3 w_i)}\; \forall\, 1\leqslant i < n \mcomma \qquad 
r_i \equiv \frac{2k}{\HH_{i-1|i} (1+3 w_i)} \; \forall\, 1<i\leqslant n \mperiod
\end{align}
\end{widetext}
In these expressions, $k$ is a fixed comoving wave mode. We next define the primordial power spectrum $\PP_{T}\supsc{end}$ and dimensionless `energy density' $\Omega_{\text{gw}}\supsc{end}$\footnote{This quantity cannot be interpreted as a real energy density, as the GW length scales are super-Hubble at that time. This improper terminology is only used based on the similarity of mathematical definitions.}, assuming that $k^{-1}$ is larger than the Hubble scale at the end of inflation by, as
\begin{align}
    \avg{h_{ij,\text{end}}(k)h^{*}_{ij,\text{end}}(k')} &= \frac{2\pi^2}{k^3}\PP_{T}\supsc{end}\delta^{(3)}(k-k') \mcomma \\
     \avg{h'_{ij,\text{end}}(k)h^{\prime*}_{ij,\text{end}}(k')} &= \frac{24\pi^2}{k^3}\HH\subsc{end}\Omega\ssgw\supsc{end}\delta^{(3)}(k-k')\mperiod
\end{align}
Finally, we consider that the present time belongs to the $n$-th era, and that at this time the GWs are sub-Hubble. In this case their energy density is well defined and we have shown that it evaluates to
\begin{align}
\label{eq:ultimate-transition}
    \nonumber 
    h^2\Omega\ssgw\supsc{today}(k) &= \left(\frac{6.8 \times 10^{20}}{1+z\subsc{end}}\right)^4\fiducial{H\ssend}{1}{\GeV}{2}\frac{1}{24}\times \\
    &\hspace{-1cm}\left\Vert T_{n-1}\cdots T_1 M_1[r_1]^{-1} 
    \begin{pmatrix}
(k/\HH\ssend) \sqrt{\PP_{T}\supsc{end}} \\[2mm]
\sqrt{12\Om\ssgw\supsc{end}}+\sqrt{\PP\supsc{end}_T}
    \end{pmatrix}\right\Vert^2 \mcomma
\end{align}
with $H_0\equiv 100 h\rm{km}/s/\rm{Mpc}$ and $r_1 = 2k/(\HH\ssend(1+3 w_1))$.

In the simplest case where perturbations are primordially frozen ($\Omega\ssgw\supsc{end}\simeq 0$) and reenter the Hubble radius during an era with equation of state $w_0$, \Eq\eqref{eq:ultimate-transition} approximates to the well known result \cite{Cai:2019cdl}
\begin{equation}
    \Om\ssgw\supsc{today}(k) \propto k^{2(3w_0-1)/(3w_0+1)} \times \PP_T\supsc{ini}(k) \mperiod
\end{equation}
Hence if $\PP_T\supsc{ini}$ is scale invariant, $\Om\ssgw\supsc{today}(k) \propto k^{-2}$ for scales reentering the Hubble radius in the matter era whereas it scales like $k^{0}$ for scales reentering during the radiation era.

\section{General ansatz for the anisotropic stress tensor}
\label{sec:ansatz}

The spin-2 anisotropic stress tensor of \Eq\eqref{eq:quadratic-Pi} in Fourier space  takes the form
\begin{align}
   \Pi^{i}_j(\kb,\tau) &=  \La^{i\, m}_{j\,\ell}(\hat{\kb})T^\ell_m(\kb,\tau)\mcomma  \\
   \label{eq:def-Tij}
T^i_j(\kb,\tau) &= C\int\hspace*{-1mm} \frac{\dd^3 \pb}{(2\pi)^3}\, \Psi^i(\pb,\tau)\Psi_j(\pb-\kb,\tau)+\dots \mcomma
\end{align}
where $\Psi_i$ \footnote{In many cases, an appropriate choice of gauge makes spatial indices sufficient to span all the degrees of freedom of the bosonic field, hence our choice of a latin index.} can generically involve components, gradients and/or time derivatives of a bosonic field of arbitrary spin with mass $m< H$.\footnote{If $m> H$ the field cannot be sizeably amplified by the expansion. We limit ourselves to bosons as Fermi exclusion complicates the amplification process in the fermionic case \cite{Adshead:2015kza,Adshead:2019aac}.} $C$ is a constant, and dots refer to traces which are removed by $\La^{i\, m}_{j\,\ell}(\hat{\kb} = \kb / |\kb|)$, the projector onto the TT part of $T^i_j$.
We assume the unequal time correlator of $\Psi_i$ to have the generic form
\begin{equation}
\label{eq:ansatz-unconstrained}
P_{\Psi}(k,\tau,\tau') = \frac{k^{n-3}M^d}{a(\tau)^ba(\tau')^b} \mcomma
\end{equation}
where $M$ denotes some arbitrary energy scale. We stress that if $\Psi_i$ involve gradients, the scaling exponent $n$ may differ from usual conventions. For instance if $\Psi_i = \partial_i\Phi$ and $\PP_\Phi\propto k^{n'}$ then $n=n'+2$, e.g., a Harrison-Zeldovich spectrum $n'=0$ for $\Phi$ actually corresponds to $n=2$ for sourcing GWs. 

The exponents $n,b,d$ are not arbitrary, as they must obey the following conditions:
\begin{enumerate}[label={\it (\roman*)}]
    \item In order for $\Pi_{ij}$ to have the correct mass dimension $4$ we must require that $\Psi_i$ has dimension $[\Psi_i] =(4-[C])/2$ in real space, that is $[\PP_{\Psi}] = 4-[C]$, i.e.,
    \begin{equation}
    \label{eq:cond-1-dim}
        n+d = 4-[C] \mperiod
    \end{equation}
    \item From Noether's theorem $T^\mu_\nu$ is always a true physical quantity \cite{maggiore_vol2}, so that itself and $\PP_\Psi$ shall not depend on the normalization of the scale factor. Under a renormalization, $a\ra Na$, the comoving momentum $k$ also scales as $k\ra Nk$. For $\PP_{\Psi}$ to remain unchanged, we must require 
    \begin{equation}
    \label{eq:cond-2-scale}
        n-2b=0 \mperiod
    \end{equation}
\end{enumerate}
This leaves $n$ as the only free exponent and \Eq\eqref{eq:ansatz-unconstrained} reduces to
\begin{equation}
\label{eq:ansatz-constrained}
    P_{\Psi}(k,\tau,\tau') = \frac{k^{n-3}}{a(\tau)^{n/2}a(\tau')^{n/2}}M^{4-[C]-n} \mperiod
\end{equation}

Applying Wick's theorem to the four-field correlator of $\Psi_i$ that we assume to be Gaussian, we obtain the unequal time correlator of $\Pi_{ij}$, 
\begin{align}
    P_\Pi(k,\tau,\tau') &= C^2\!\!\int\hspace{-1mm} \frac{\dd^3\pb}{(2\pi)^3} P_{\Psi}(|\pb|,\tau,\tau') \times \qquad\qquad \nonumber \\
\label{eq:Psi-correlator-integral}
    & \qquad\quad  P_{\Psi}(|\kb-\pb|,\tau,\tau')f(\theta,\varphi)\mperiod
\end{align}
Here $f(\theta,\varphi)$ is a function of the spherical angles coming from the projection tensor $\La^{i\, m}_{j\,\ell}(\hat{\kb})$ which affects the amplitude by a factor of order unity and which we ignore in what follows. As our main focus is the scale dependence of $\PP_T$, we are less careful in computing its amplitude and often neglect dimensionless numerical factors.

If $P_{\Psi}(p)$ is a pure power law with spectral index $n>0$, the above integral is apparently divergent in the ultraviolet (UV). This well known UV divergence must be subtracted. This can be done in different ways, e.g., via adiabatic regularization~\cite{Parker:1974qw,Animali:2022lig,Kolb:2023ydq}. Although a more formal renormalization procedure is in principle required, the key physical input for adiabatic regularization is that modes that oscillate fast are not amplified. Only large enough scales which have already exited the Hubble radius become sizeably populated during inflation and can source tensor metric perturbations, while sub-Hubble vacuum fluctuations do not. They are in fact still in the asymptotically flat Minkowski vacuum for which the formally divergent amount of GWs is known not to contribute to observables.  As a simple approximation, we shall thus truncate the scales responsible for GW production around the Hubble radius, $\cut^{-1} \simeq \HH^{-1}\simeq \tau/q$. This scaling with the Hubble parameter is found in treatments of UV regularization in inflation~\cite{Zeldovich:1971mw,Marozzi:2011da, Animali:2022lig, Kolb:2023ydq}. Again, despite being obviously less accurate than a proper renormalization scheme, it is natural that the integral \eqref{eq:Psi-correlator-integral} is governed by the expansion rate, as it must vanish in the flat space case where no amplification of the vacuum fluctuations occurs. Besides, studies that have tackled the notoriously difficult task of computing one-loop corrections to the power spectrum of inflationary perturbations do find, up to usual logarithmic corrections in the renormalization scale, that the Hubble parameter is indeed the relevant energy scale \cite{Animali:2022lig,Galanti:2024jhw,Braglia:2025qrb}.

We therefore study the super-Hubble regime $k\ll \HH \simeq \cut$. We first approximate\footnote{We perform a more careful treatment of this integral in \cite{Teuscher:2025jhq}.}
\begin{equation}
\label{eq:PPI-split-in-two}
    P_\Pi(k) \simeq \int_0^k \hspace{-2mm} \dd p\, p^2 P_{\Psi}(p)P_{\Psi}(k) + \!\int_k^{\La_{\text{\tiny{UV}}}}\hspace{-4mm} \dd p\, p^2 P_{\Psi}(p)P_{\Psi}(p) \mcomma
\end{equation}
which leads us to scrutinize three different cases (neglecting logarithmic corrections): 

\begin{enumerate}[label={\it (\roman*)}]
\item $n\geqslant 3/2$:\ $P_\Pi$ is dominated by the UV cutoff and has a white noise spectrum, 
\begin{equation}
    P_\Pi(k) \propto \bigO(k^{2n-3})+\bigO(\cut^{2n-3}) \propto \cut^{2n-3}\mperiod
\end{equation}
Note that here $P_\Pi(k)$ is not determined by the source spectrum at scale $k$ but by its UV cutoff. This `sweeping' of power from small to larger scales is a common phenomenon in nonlinear processes, which can also take place in loop corrections, see, e.g., Ref.~\cite{Iacconi:2023ggt}.
\item $3/2> n\geqslant 0$: \ $P_\Pi$ is a combination of two terms dominated by $k$, and as such
\begin{equation}
P_\Pi(k) \propto \bigO(k^{2n-3})+\bigO(\cut^{2n-3})\propto k^{2n-3} \mperiod
\end{equation}
In both cases, although the convolution in \Eq\eqref{eq:Psi-correlator-integral} introduces a coupling between different modes \cite{Iacconi:2023ggt}, the strong hierarchy between the scales in the problem makes the approximate expression of $P_\Pi$ fairly concise. 
\item $n<0$: \  an infrared cutoff $\cutir$ must be introduced for the integral \eqref{eq:Psi-correlator-integral} to converge. This could, for example, be the largest scale that exited the Hubble radius during inflation. For $k\gg \cutir$, $P_\Pi$ is then given by 
\begin{equation}
P_\Pi(k) \propto k^{n-3}\cutir^n\mperiod 
\end{equation}
Properly treating this case would in principle require a consistent IR regularization as, for example, proposed in~\cite{Marozzi:2011da}. Contrary to the UV divergence it may be that in this case the regulated result departs significantly from our approximation, but we shall not study this case in detail.
\end{enumerate}

To conclude this analysis, we observe that in the computation of $P_\Pi(k,\tau,\tau')$ the UV cutoff $\cut(\tau)\simeq\HH(\tau)$ may in principle depend both on $\tau$ and $\tau'$. Given the prior discussion on fluctuations in an asymptotically flat space, we shall set
\begin{align}
    \cut(\ta,\ta') &= \mathcal{O}(1)\times\min (\HH(\tau),\HH(\tau')) \nonumber \\
\label{eq:cutoff-def-with-max}    
    &= \frac{\mathcal{O}(1)}{\max(|\tau|,|\tau'|)} \mcomma
\end{align}
effectively excluding modes that are sub-Hubble {\it at least} at one of the two times. This provides an explicit shape for the kernel $K$ defined below \Eq\eqref{eq:PT-formula}. In the Appendix we show that this kernel meets the aforementioned positivity criterion, ensuring the consistency of that choice.

\section{Universal scaling of the tensor power spectrum}
\label{sec:universal}


We now combine the previous results to obtain \Eq\eqref{eq:nT-main-formula}. Let us first consider the case $n\geqslant 3/2$, as typical sources fall into this category (see examples below).  Inserting $P_\Pi$ into \Eq\eqref{eq:PT-formula}  yields the following super-Hubble tensor power spectrum 
\begin{widetext} 
\begin{equation}
\PP_T(k,\tau) \simeq \frac{C^2}{2\pi^2}\frac{M^{4-2(n-2)-2[C]} H\subsc{end}^{2q(2-n)}}{\planckmass^4}\left(\frac{k}{a\subsc{end}}\right)^{2(n-2)(q+1)} 
\hspace{-1mm} \int_{k\tau_i}^{k\tau} \frac{\dd y}{y}\int_{k\tau_i}^{k\tau} \frac{\dd z}{z} \frac{(yz)^{q(2-n)+2}}{|\max(-y,-z)|^{2n-3}} \mperiod 
\label{eq:Ph-res1}    
\end{equation}
\end{widetext} 
The label `end' refers to some fixed time point, in general the end of inflation. 

\begin{figure}
    \centering
    \includegraphics[width=\linewidth]{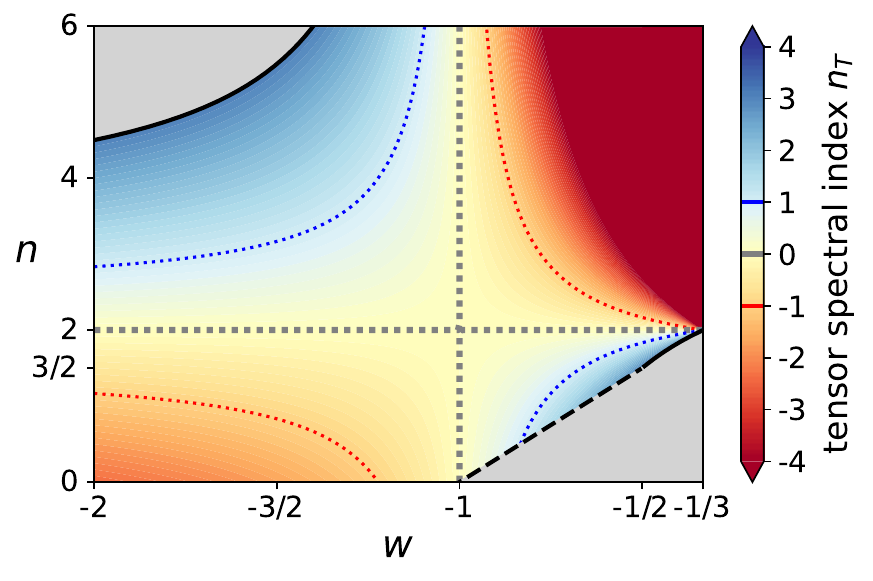}
    \caption{GW spectral index $n_T$ from \Eq\eqref{eq:nT-main-formula}. For proper visualization, all $n_T<-4$ are displayed with the same color. Dotted lines indicate $n_T=-1,0,1$. Solid (resp. dashed) black lines mark out regions where the inequality $2n(1+w)\geqslant 5+7w$ (resp. $n\geqslant3(1+w)$) fails. As an example, sources corresponding to $n=2$ lead to a nearly scale invariant SGWB for all $w<-1/3$. For completeness we have included the region $w<-1$ where our analysis holds, although this case is somewhat  exotic.}
    \label{fig:plot-n-w}
\end{figure}

In the integral \eqref{eq:Ph-res1}, $\tau_i$ corresponds to the time $\tau_k = -\mathcal{O}(1) / k$ when the mode $k$ exits the Hubble radius, i.e. $x_i = k\tau_i= k \tau_k = -\mathcal{O}(1) \propto k^0$ is, crucially, {\it independent of $\mathrm{k}$}. Provided that this integral is dominated at its $k$-independent lower boundary $x_i$, it is then only a numerical value that has no effect on the spectral index of $\PP_T$ -- and the exact, model-dependent expression of $x_i$ becomes irrelevant. This happens to be the case if the condition 
\begin{equation}
    \label{eq:technical-condition}
    2n(1+w) \geqslant 5+7w
\end{equation}
 is fulfilled,\footnote{The condition $n\geqslant 3(1+w)$ is also necessary, but redundant, since for $n\geqslant 3/2$, $2n(1+w)\geqslant5+7w$ implies $n\geqslant3(1+w)$.} since $\left|x_i\right|\gg \left|x\right|$. We therefore obtain
\begin{equation} 
\label{eq:Ph-res2}
\PP_T(k,\tau) \propto \frac{C^2}
{2\pi^2}\frac{M^{4-2(n-2)-2[C]} H\subsc{end}^{2q(2-n)}}{\planckmass^4}k^{2(n-2)(q+1)} \mcomma
\end{equation}
establishing \Eq\eqref{eq:nT-main-formula}. One can check that this result holds also for $0\leqslant n < 3/2$, if the requirement $n\geqslant 3(1+w)$ is satisfied instead. These two validity conditions are very easily met and span most of the physically relevant part of the parameter space, as illustrated in Figure \ref{fig:plot-n-w}.

Let us pinpoint that a consistency issue seemingly arises from the use of \Eq\eqref{eq:approx-green-func}, if the integral \eqref{eq:Ph-res1} runs up to $|x_i|=\mathcal{O}(1)$. Although close to $\left|y\right|=1$ or $\left|z\right|=1$ the use of \Eq\eqref{eq:mildly-approx-green-func} is more appropriate, it is clear from \Eq\eqref{eq:Ph-res1} that this replacement will only affect the overall amplitude of $\PP_T$ rather than its scale dependence, providing $x_i$ still dominates the integral. This is the case whenever condition \eqref{eq:technical-condition} is satisfied. The value of $n_T=2(n-2)(q+1)$ remains therefore unaffected.

For sources with red power spectra, $n<0$, we only quote the result for $-1/3 > w \geqslant -1$. We treat $\cutir$ as time independent, as it can typically represent the value $-\tau\subsc{in}\simeq \cutir^{-1}$ at the beginning of inflation. For $k\gg\cutir$,
the GW power spectrum is not frozen anymore and its value at the end of inflation becomes
\begin{equation}
    \PP_T(k,\tau\subsc{end}) \propto \frac{C^2}{2\pi^2}\frac{M^{4-2(n-2)-2[C]} }{\planckmass^4 H\subsc{end}^4} \left(\frac{k \cutir}{a\subsc{end}^2}\right)^n\mperiod
    \label{eq:Ph-res3}   
\end{equation}
This GW spectrum is identical to the spectrum of $\Psi_i$, but multiplied by a factor of $(\cutir/\HH\subsc{end})^n / \planckmass^4$ (taking $M=H\subsc{end}$ and $C=\mathcal{O}(1)$ here).
Note that as $n<0$ this is typically an enormous enhancement factor that virtually rules out such models. In slow-roll, where $q\simeq -1$, the enhancement with respect to the corresponding model with nonnegative spectral index \eqref{eq:Ph-res2} is
\begin{align}
&\left.\PP_T(k,\tau\subsc{end})\right|\subsc{IR} = D(k) \left.\PP_T(k)\right|\subsc{UV} \\
 \text{with} \quad&
 D(k) = \exp[-n(N_k + N\subsc{tot})]\gg 1\mcomma
\end{align}
where $N_k$ is the number of $e$-folds of inflation after the Hubble exit of the mode $k$, and $N\subsc{tot}$ is the total number of $e$-folds. As mentioned earlier, we emphasize that a careful procedure of IR regularization may nonetheless lead to a result quite different from the prediction involving a simple IR cutoff \cite{Marozzi:2011da}.

Let us now return to our main result which consists of the more common case $n\geqslant 0$ where no IR regularization is needed. We first point out that the time integral \eqref{eq:Ph-res1} being dominated by early times indicates that most of GW production on scale $k$ happens right after Hubble crossing. This was also argued in Ref.\cite{Caprini:2014mja}. As a consequence, the spectrum \eqref{eq:Ph-res2} is frozen. 

In addition, \Eq\eqref{eq:Ph-res2} shows that, regardless of the source spectral index, during slow-roll $q\simeq -1$ the GW spectrum is always (nearly) scale invariant. We attribute this property to a balance between dilution (due to the expansion of the Universe) and power scaling. Even though, contrary to scale invariant ones, blue and white noise spectra have more power on smaller scales, these scales exit the Hubble radius later than larger ones. As GW production cannot happen before Hubble crossing, the available power has been more severely damped by expansion. The dimensional constraints used to establish the ansatz \eqref{eq:ansatz-constrained} show that these scalings are actually intertwined (bluer spectra dilute faster), and so the cancellation happens for all $n\geqslant 0$, leading to a universal scale invariance of the SGWB induced during slow-roll inflation.

On the other hand, if $q$ deviates significantly from $-1$ we observe that only the case $n\simeq 2$ leads to a scale invariant SGWB, yet irrespective of the background equation of state $w$. If $\Psi_i$ is the gradient of some bosonic field, $\Psi_i = \dr_i\Phi$, this requires the spectrum of $\Phi$ to be itself scale invariant, $n_\Phi \simeq 0$. This is precisely the case for second-order scalar perturbations (see details below) for which the induced GW spectrum is therefore always nearly scale invariant.
Moreover, if $w>-1$, hence $q+1<0$, if the underlying field has a blue spectrum, $n>2$, the induced SGWB is red and vice versa, a red source spectrum $n<2$ induces a blue GW spectrum, see Figure \ref{fig:plot-n-w}. 

Finally, if the only scales in the problem are the Planck mass and the Hubble scale, we find the concise relation
\begin{equation}
    \label{eq:Ph-res4}
\hspace*{-3mm}\PP_T \propto \frac{H\subsc{end}^{4}}{\planckmass^4}\left(\frac{k}{\HH\subsc{end}}\right)^{n_T} \hspace{-3mm}, \quad n_T=2(n-2)(q+1) \mperiod
\end{equation} 


\section{Applications}
\label{sec:appli}

Let us illustrate our general prediction \eqref{eq:Ph-res2} for two specific examples of driven GW production.

\bigskip 

\subsection{Scalar induced GWs} 

Scalar induced GWs have been widely discussed (see Ref. \cite{Domenech:2021ztg} for a review) as a natural GW production mechanism during inflation. Although the full picture is of course more complicated, a gauge-invariant scalar perturbation $\Phi$ typically generates a stress tensor (removing isotropic components) \cite{Durrer:2020fza, Baumann:2007zm}
\begin{equation}
    T_{ij} \sim \frac{4 \planckmass^2}{3(1+w) a^2}\partial_i\left(\Phi+\frac{\Phi'}{\HH}\right)\, \partial_j \left(\Phi+\frac{\Phi'}{\HH}\right) \mcomma
\end{equation}
so $\Psi_i = (1/a)\partial_i(\Phi+\Phi'/\HH)$ and $C=4 \planckmass^2 /(3(1+w))$ here. Assuming slow-roll, the curvature perturbation $\zeta$ has
a flat Harrison-Zeldovich spectrum $\PP_\zeta \simeq H^2/\left(8\pi^2\epsilon \planckmass^2\right)$ \cite{maggiore_vol2}. It is linked to $\Psi_i$ by $\PP_\Psi \simeq (k/a)^2 \ep^2 \PP_{\zeta}$ \cite{Durrer:2020fza}, using $C\simeq 2 \planckmass^2/\epsilon$. This matches the ansatz \eqref{eq:ansatz-constrained} with $n=2$ and an additional dimensionless factor $\ep H^2/(8\pi^2 \planckmass^2)$ that can be reabsorbed into $C$.
\Eq\eqref{eq:Ph-res2} thus predicts (at lowest order in slow-roll) a scale invariant spectrum of scalar-induced GWs, in accordance with Refs. \cite{Baumann:2007zm,Domenech:2021ztg}.

Moreover, even beyond slow-roll, scalar fields can generate a scale invariant spectrum of GW if they are nonminimally coupled to gravity. One may, e.g., consider adding a coupling $\xi R\phi^2$ ($R$ is the Ricci scalar) to the Lagrangian of a scalar field $\phi=\bar{\phi}+\delta\phi$, modifying the evolution of first-order perturbations. Using a gauge invariant combination $\delta\phi\subsc{g.i.}\simeq (\bar{\phi}'/\HH)\zeta$ \cite{Durrer:2020fza}, one shows that for $\xi = \frac{1}{6}-\frac{1}{3q(q-1)}$ one recovers $\PP_{\Phi+\Phi'/\HH}\propto k^0$ hence $n=2$. 

In the literature (see, e.g., Refs. \cite{Baumann:2007zm,Domenech:2021ztg}) it has been shown that this type of secondary induced GWs can overcome first order inflationary GWs if and only if the latter lead to a tensor-to-scalar ratio $r<10^{-6}$. Only in this case can the former GWs be unambiguously detected. But even then, their detection would be very ambitious, and no experiment is presently foreseen that could measure this `guaranteed' SGWB.



\subsection{Gauge field induced GWs}
Another well-studied mechanism \cite{Durrer:2023rhc,Teuscher:2025jhq} of inflationary GW production is through the existence of primordial electromagnetic fields, amplified out of the vacuum by nonconformal couplings like in the Ratra model \cite{ratramodel}. In this case their stress-energy tensor reads \cite{jackson2021classical}
\begin{equation}
    T_{ij} = E_i E_j + B_i B_j -\frac{1}{2}\delta_{ij}(E^2+B^2)\mcomma
\end{equation}
where $E_i, B_i$ are the electric and magnetic fields. Focusing on the latter this would simply correspond to $\Psi_i = B_i$. The exact scaling of $\Psi_i$ then depends on the model considered, but it is typically very blue \cite{Teuscher:2025jhq,Barnaby:2011vw} ($n=4$); in that case a scale invariant GW spectrum is only expected for $w\simeq-1$. For almost De Sitter $q=-1-\epsilon$ we expect from \Eq\eqref{eq:Ph-res2} $\PP_T \propto k^{-4\epsilon}$, agreeing -- see details below -- with Refs. \cite{Teuscher:2025jhq, Caprini:2018mtu, Bartolo:2016ami}. Here the predicted spectral tilt differs from the standard inflationary case $\PP_T\propto k^{-2\epsilon}$.

Despite these agreements with previous studies, we show here that model-dependent corrections to the tilt obtained in \Eq\eqref{eq:Ph-res2} can arise. The principal reason is that the constant $C$ may itself slowly drift in more realistic setups. As an example, let us consider the widely discussed case of axion inflation \cite{Pajer:2013fsa}. In this scenario, the pseudoscalar inflaton field $\phi$ is coupled to an Abelian gauge field via 
\begin{eqnarray}
    \frac{\alpha}{f_a}\phi F_{\mu\nu}\tilde{F}^{\mu\nu}\mcomma
\end{eqnarray}
where ($\tilde{F}_{\mu\nu}$) $F_{\mu\nu}$ is the (dual) electromagnetic tensor, $f_a$ has the dimension of a mass and $\alpha$ parametrizes the coupling strength. Like the Ratra model illustrated above, gauge fields amplified via this coupling in turn generate GWs, whose power spectrum is generally found to be \cite{Caprini:2018mtu, Barnaby:2011vw}
\begin{eqnarray}
    \PP_T \propto \frac{1}{2\pi^2}\left(\frac{H}{\planckmass}\right)^4 \frac{e^{4\pi\xi}}{\xi^6}\mperiod
\end{eqnarray}
The parameter $\xi \equiv \alpha\phi'/(2 f_a \HH)$ governs the field amplification, and $\xi \propto \sqrt{\ep}$ during slow-roll. Compared to the standard inflationary SGWB $\PP_T \propto(H/\planckmass)^2$ this presents an additional scaling with $(H/\planckmass)^2$, which can however be overcome by the prefactor $\exp(4\pi\xi)/\xi^6$ that can be significant, so that this SGWB may be observable, see Ref.~\cite{Teuscher:2025jhq} for more details.


As mentioned earlier, at next-to-leading order in $\ep$ this matches our prediction \eqref{eq:Ph-res2} by identifying $n_T=-4\ep$ and $C=e^{2\pi\xi}/\xi^3$. However, across long time periods (or equivalently long range of scales) one cannot ignore that the value of $\xi$ evolves (as inflation must end). Evaluating $C$ at the Hubble crossing time results in \cite{Caprini:2018mtu}
\begin{equation}
    n_T = -4\ep +(4\pi\xi-6)(\ep-\eta)\mcomma
\end{equation}

where $\eta\equiv -\ddot{\phi}/(H\dot{\phi}) \ll 1$ is the second    slow-roll parameter. This also amounts to saying that the value of $w$ (or $\ep$) across sufficiently many $e$-folds cannot be held constant, and therefore \Eq\eqref{eq:nT-main-formula} only applies to a more limited range of scales \cite{Domcke:2016bkh}.

Nonetheless, we emphasize that these corrections to the tilt do not spoil the qualitative aspect of the results presented above, as they will generically be proportional slow-roll parameters. Either $n_T$ remains close to near scale invariance, or \Eq \eqref{eq:nT-main-formula} predicts a value of $n_T$ far from zero, in which case these corrections will be of subleading order. 

Furthermore, in this and probably most examples, our law for the spectral index can suffer from  significant corrections when applied to the smallest scales, that are amplified close to the end of inflation. On these scales, the source may not have redshifted away at the start of reheating and in the subsequent radiation era when the equation of state is indeed changing significantly. However, for cosmologically relevant scales that exit the Hubble radius long before the end of inflation, we expect corrections to be small. Even if the source may still be present after inflation, it will be strongly redshifted. In the example of electromagnetic fields, the electric field is damped exponentially by the high conductivity of the cosmic plasma, while the magnetic field is also damped on small scales that are the most relevant sources of GWs during inflation. Therefore, even though the magnetic field on large scales survives until today, it only contributes a subleading SGWB.

\section{Conclusion}
\label{sec:conclu}

In this paper we have exposed, using dimensional arguments, how induced gravitational wave backgrounds acquire a spectral tilt whose value is uniquely determined by the spectral index of their source and by the inflationary dynamics. During slow-roll inflation the generated background is systematically nearly scale invariant with a small deviation given by the slow-roll parameter $\ep$. This is the main result, outlined in \Eq\eqref{eq:nT-main-formula}. To summarize, let us recall the assumptions leading to this relation: {\it (i)} we consider an inflationary phase with either constant equation of state $w<-1/3$ or slow-roll; {\it (ii)} only super-Hubble perturbations generate GWs, as is the case if the source field $\Psi_i$ was amplified from the vacuum; {\it (iii)} $\Psi_i$ is a Gaussian field; {\it (iv)} its spectrum -- that can be approximated by a generic power law of $k$, of $a$, and of an arbitrary energy scale
-- is not red, $n\geqslant 0$; {\it (v)} a technical but not very restrictive inequality between $n$ and $w$ must be satisfied -- see Figure \ref{fig:plot-n-w}. 

Would assumption {\it (i)}, {\it (ii)} or {\it (iv)} be relaxed, our derivation may however still proceed, as long as the dependence on $k$ of  the bounds dominating integrals \eqref{eq:Psi-correlator-integral} and \eqref{eq:Ph-res1} is carefully established. Condition {\it (ii)} can easily be generalized to broken power laws. The case of peaked spectra has also been discussed in the literature \cite{Pi:2020otn}, but in general one can expect peaked and blue spectra to contribute similarly to \Eq\eqref{eq:Psi-correlator-integral} if the peak is located around the Hubble scale. Other works dedicated to SGWBs generated during the radiation era have, e.g., considered GW generation time scales to be independent of $k$, in which case $x_i=k\tau_i\propto k^1$ and \Eq\eqref{eq:Ph-res1} leads to $\PP_T\propto k^3$, in full agreement with Ref. \cite{Cai:2019cdl}. Let us also comment that the super-Hubble sources of anisotropic stress discussed in this paper may continue to exist after the transition from inflation to the radiation era and to participate in the overall SGWB. Nevertheless, the dilution of their stress energy tensor with the expansion should render their contribution much smaller than the contribution from inflation that is mainly generated at the time they exit the Hubble scale.


Under the previous assumptions, we have shown that the source field generates a stochastic GW background with spectral index $n_T=6(n-2)(1+w)/(1+3w)$. Noticeably, if $w\simeq -1$ it is nearly scale invariant independently of $n$. There can be additional slow-roll corrections to the tilt if considered over very large ranges of scales \cite{Domcke:2016bkh}. Furthermore, if the only scales in the problem are the Hubble and Planck scales, the spectrum amplitude is proportional to $H^{4}/\planckmass^4$. Naively, this factor is an additional $(H/\planckmass)^2$  suppression with respect to the direct, quantum GWs background. Therefore to make such a background visible, we either need an additional enhancement, like for axion inflation, see above, or a very large number of fields that contribute, see Ref. \cite{delRio:2018vrj}. In general  sizable, model-dependent corrections to the amplitude are expected \cite{Teuscher:2025jhq, Caprini:2018mtu}, and we cannot provide an immediate general comparison with observations.

The goal of our prediction is therefore not to give an exact, model specific value of the induced GW spectral tilt, but rather to serve as a fast, efficient way to estimate it for any effective model of anisotropic stress.

Although a detailed discussion is beyond the scope of this work, we further expect the stochastic GWs sourced in this way to exhibit non-Gaussian statistics, as they are generated from terms quadratic in the Gaussian source field. In future work we hope to carry the approach that we developed here to higher order statistical observables.


\bigskip

\noindent{\bf Acknowledgments:~~} The authors thank Misao Sasaki and Valerie Domcke for useful exchanges. M.T. thanks the IDEX for funding his mobility grant.

\appendix

\section{Appendix: proof that \intitle{$(u,v)\mapsto \max(u,v)^{-1}$} is a positive kernel}
\label{appx:kernel}
\noindent A positive
kernel is a symmetric function $K:\R_+^2\to\R$ satisfying (among other requirements) the Mercer condition \cite{Mercer:1909dea}, namely: for any $I\subset \R_+$ and $f\in L^2(I)$, $\int\!\!\!\int_{I^2} f(y)f(z)K(y,z)\dd y\dd z \geqslant 0$. We show here the following result:

{\it For any decreasing (resp. increasing) function $\phi:\R_+\to\R_+$, $K = \phi \circ \max$ (resp. $K = \phi\circ \min$) is a positive kernel.}

The choice of function $\phi:u\mapsto 1/u$ made for $\cut$ in \Eq \eqref{eq:cutoff-def-with-max} thus ensures that $\PP_T$ given by \Eq\eqref{eq:PT-formula} is always a positive quantity.

{\it Proof:} Given $y,z\in\R_+$, we write $\phi(\max(y,z)) = \int_0^\infty \mathbbm{1}_{\{s\leqslant \phi(\max(y,z))\}}\dd s$, with $\mathbbm{1}$ denoting the indicator function. As $\phi$ is monotonically decreasing, $\{s \leqslant \phi(\max(y,z))\} = \{s \leqslant \phi(y)\}\cap \{s \leqslant \phi(z)\}$. Thus, after a permutation,
\begin{eqnarray}
\nonumber
 \int\!\!\!\int_{I^2} f(y)f(z)K(y,z)\dd y\dd z &=& \\
      && \hspace{-4cm} \int_0^\infty \dd s \left(\int_I f(y) \mathbbm{1}_{\{s\leqslant \phi(y)\}}\dd y\right)^2 \geqslant 0\mperiod
\end{eqnarray} 
A similar proof follows in the case where $\phi$ is increasing.

\bibliographystyle{apsrev4-2}
\bibliography{references}

\end{document}